\documentclass[%
 aip,
 amsmath,amssymb,
 reprint,%
]{revtex4-1}

\usepackage{graphicx}
\usepackage{dcolumn}
\usepackage{bm}

\usepackage[utf8]{inputenc}
\usepackage[T1]{fontenc}
\usepackage{mathptmx}
\usepackage{hyperref}
\usepackage[capitalise]{cleveref}
\begin{document}

\newcommand{\sm}[1]{{\scriptscriptstyle#1}}

\preprint{AIP/123-QED}

\title[Kinetic modelling of carrier cooling in lead halide perovskite materials]{Kinetic modelling of carrier cooling in lead-halide perovskite materials}

\author{Thomas R. Hopper}
\thanks{These authors contributed equally to the manuscript}
\affiliation{Ultrafast Optoelectronics Group, Department of Chemistry, Imperial College London, London W12 0BZ, United Kingdom}
\author{Ahhyun Jeong}
\thanks{These authors contributed equally to the manuscript}
\affiliation{Ultrafast Optoelectronics Group, Department of Chemistry, Imperial College London, London W12 0BZ, United Kingdom}
\author{Andrei Gorodetsky}
\affiliation{Ultrafast Optoelectronics Group, Department of Chemistry, Imperial College London, London W12 0BZ, United Kingdom}
\author{Franziska Krieg}
\affiliation{Laboratory of Inorganic Chemistry, Department of Chemistry and Applied Biosciences, ETH Zurich, Vladimir-Prelog-Weg 1-5/10, 8093 Zurich, Switzerland}
\affiliation{Laboratory for Thin Films and Photovoltaics, Empa – Swiss Federal Laboratories for Materials Science and Technology, Überlandstrasse 129, CH-8600 Dübendorf, Switzerland}
\author{Maryna I. Bodnarchuk}
\affiliation{Laboratory of Inorganic Chemistry, Department of Chemistry and Applied Biosciences, ETH Zurich, Vladimir-Prelog-Weg 1-5/10, 8093 Zurich, Switzerland}
\affiliation{Laboratory for Thin Films and Photovoltaics, Empa – Swiss Federal Laboratories for Materials Science and Technology, Überlandstrasse 129, CH-8600 Dübendorf, Switzerland}
\author{Xiaokun Huang}
\affiliation{Institute for High-Frequency Technology, Technische Universität Braunschweig, Schleinitzstrasse 22, 38106, Braunschweig, Germany}
\author{Robert Lovrincic}
\affiliation{Institute for High-Frequency Technology, Technische Universität Braunschweig, Schleinitzstrasse 22, 38106, Braunschweig, Germany}
\author{Maksym V. Kovalenko}
\affiliation{Laboratory of Inorganic Chemistry, Department of Chemistry and Applied Biosciences, ETH Zurich, Vladimir-Prelog-Weg 1-5/10, 8093 Zurich, Switzerland}
\affiliation{Laboratory for Thin Films and Photovoltaics, Empa – Swiss Federal Laboratories for Materials Science and Technology, Überlandstrasse 129, CH-8600 Dübendorf, Switzerland}
\author{Artem A. Bakulin}
\thanks{Corresponding author email: a.bakulin@imperial.ac.uk}
\affiliation{Ultrafast Optoelectronics Group, Department of Chemistry, Imperial College London, London W12 0BZ, United Kingdom}

\begin{abstract}
The relaxation of high-energy “hot” carriers in semiconductors is known to involve the redistribution of energy between (i) hot and cold carriers and (ii) hot carriers and phonons. Over the past few years, these two processes have been identified in lead-halide perovskites (LHPs) using ultrafast pump-probe experiments, but the interplay between these processes is not fully understood. Here we present a comprehensive kinetic model to elucidate the individual effects of the hot and cold carriers in bulk and nanocrystal CsPbBr\textsubscript{3} films obtained from “pump-push-probe” measurements. In accordance with our previous work, we observe that the cooling dynamics in the materials decelerate as the number of hot carriers increases, which we explain through a “hot-phonon bottleneck” mechanism. On the other hand, as the number of cold carriers increases, we observe an acceleration of the cooling kinetics in the samples. We describe the interplay of these opposing effects using our model, and by using series of natural approximations, reduce this model to a simple form containing terms for the carrier-carrier and carrier-phonon interactions. The model can be instrumental for evaluating the details of carrier cooling and electron-phonon couplings in a broad range of LHP optoelectronic materials. 
\end{abstract}
\maketitle

\section{\label{sec:level1}Introduction}
Lead-halide perovskites (LHPs) are rapidly emerging as a promising class of semiconducting material for future optoelectronic applications. Since the first report of 3.8\% power conversion efficiency (PCE) in 2009,\cite{Kojima2009} laboratory-scale LHP solar cells can now routinely achieve PCEs above 20\%, with a record PCE of $\sim$25\%.\cite{NREL2017} Compared to the top PCEs of other third-generation solution-processable photovoltaic materials, this value is impressively close to the $\sim$33\% PCE given by the theoretical “Shockley-Queisser” limit for single-junction solar cells.\cite{Shockley1961} A key factor underpinning this limit is the rapid dissipation of energy from “hot” carriers following the photoexcitation of the semiconductor above its bandgap. Early work by Ross and Nozik postulated that the performance of a solar cell could be dramatically enhanced if the hot carriers could be utilised.\cite{Ross1982} The efficiency of this process depends on how quickly hot carriers dissipate their excess energy, reflected by the cooling rate.\\
\indent The cooling rate depends on many processes which control the carrier dynamics, including the interaction between carriers (i.e. redistribution of heat between the carriers, sometimes called thermalisation), and the interactions between carriers and phonons (i.e. the removal of excess energy through lattice vibrations).\cite{Kahmann2019} The latter process has been studied extensively in LHPs using pump-probe methods. A common finding is a slowdown of the cooling kinetics with increasing carrier density.\cite{Price2015,Yang2016,Yang2017,Papagiorgis2017,Li2017,Chen2019c,Diroll2019a} This phenomenon has also been observed in traditional semiconductors such as GaAs,\cite{Shah1978,Leheny1979} and is often referred to as the “hot-phonon bottleneck”. For LHPs, this behaviour is thought to occur when polarons (i.e. localised distortions of the lattice surrounding a carrier) spatially overlap, leading to the reabsorption of hot phonons by adjacent carriers.\cite{Frost2017a} We previously demonstrated that this behaviour in LHPs is sensitive to the material composition, and is particularly pronounced in the all-inorganic CsPbBr\textsubscript{3} system compared to its hybrid counterparts with vibrationally active organic cations.\cite{Hopper2018}
Most commercially available transient absorption methods do not have the time resolution required to observe the aforementioned thermalisation process, but high-resolution two-dimensional electronic spectroscopy has been employed to demonstrate this effect in LHPs.\cite{Richter2017,Ghosh2017} These studies show that the high-energy carriers produced by above-gap excitation reach a thermal distribution through inelastic carrier scattering at the early stage of cooling (<100 fs), followed by the carrier-phonon interactions where the heat is dissipated in the lattice ($\sim$500 fs). Although these reports provide valuable insight into the mechanisms of carrier relaxation, the methods lack the ability to distinguish the individual effects of the hot and cold carriers on carrier cooling.\\
\indent Here, we propose a comprehensive kinetic model to study the effect of the number of hot and cold carriers on the carrier cooling kinetics in CsPbBr\textsubscript{3} bulk and nanocrystal (NC) films determined by “pump-push-probe” measurements. Our results convey two competing pathways for carrier cooling based on the dissipation of the excess hot carrier energy into either optical phonons or cold carriers. By applying simple approximations, we reduce this model into a single equation containing terms for the carrier-carrier and carrier-phonon interactions. At sufficiently high cold carrier concentration, the interactions between hot and cold carriers dominate over the hot-phonon bottleneck behaviour, leading to an overall increase in the rate of carrier cooling.

\section{\label{sec:level2}Experimental methods}
\subsection{Sample preparation}
CaF\textsubscript{2} substrates (12 mm diameter, 1 mm thickness, EKSMA Optics) were first treated with O\textsubscript{2} plasma to improve wetting.

The bulk CsPbBr\textsubscript{3} films were prepared by dissolving 1:1 mole ratio of PbBr\textsubscript{2} (Alfa Aesar, 99.9\%) and CsBr (Alfa Aesar, 99.9\%) in dimethyl sulfoxide (Sigma-Aldrich, 99.9\%) with concentration of 0.4 M and stirring at 50~$^{\circ}$C overnight. Afterwards, $\sim$35~$\mu$l of the as-prepared solution was dropped onto the substrate and spin-coated at 3000~rpm for 2~min. The solvent extraction method was then applied by using 300~$\mu$l chloroform after 1~min spinning.\cite{Kerner2016,Endres2016} The as-formed thin films were then thermally annealed on a hot plate at 100~$^{\circ}$C for 10~min inside the glovebox. The thin film layer thicknesses were determined by UV-vis ellipsometry to be $\sim$300~nm.

The CsPbBr\textsubscript{3} NCs were synthesised according to a previously reported method.\cite{Krieg2018} 1~mL lead-oleate (0.5~M in 1-octadiene (ODE)), 0.8~mL Cs-oleate (0.4~M in ODE), 0.0445~g of N,N-(dimethyloctadecylammonio)-propanesulfonate and 10~mL ODE were added to a 25~mL three-neck flask and heated to 130~$^{\circ}$C under vacuum. As soon as the reaction temperature was reached, the atmosphere was changed to nitrogen and 0.42~g of oleylammonium bromide, dissolved in 3~mL of toluene, was injected. The reaction mixture was immediately cooled to room temperature by means of an ice bath. The crude solution was then centrifuged at 29500~g for 10~min. The precipitate was discarded and ethylacetate (30~mL) was added to the supernatant (ca. 15~mL). The mixture was centrifuged at 29500~g for 10~min and the precipitate was dispersed in toluene (3~mL). This solution was subjected to three additional rounds of precipitation with ethylacetate (6~mL) and centrifugation at 29500~g for 1~min and subsequent redispersion using toluene (3~mL). The film thickness was determined to be $\sim$300~nm by spectroscopic ellipsometry.

\subsection{UV-Vis spectroscopy}
In order to determine the carrier density induced by the pump, the absorption cross section of the samples was calculated from the linear absorption spectra. These spectra were obtained from a UV-Vis spectrometer (Shimadzu 2600, equipped with an ISR-2600Plus integrating sphere). A 1~nm sampling interval was used with a 5~nm slit width. 

\subsection{Ultrafast differential transmission measurements}
A Ti:sapphire regenerative amplifier (Astrella, Coherent) was used to pump two optical parametric amplifiers (TOPAS-Prime, Coherent) with 800~nm pulses at a repetition rate of 4~kHz and pulse duration of $\sim$35~fs. The signal output ($\sim$1300~nm) of one of the optical parametric amplifiers was directed into a $\beta$-barium borate crystal (EKSMA Optics) along with the residual part of the fundamental (800~nm) from the regenerative amplifier. Sum frequency generation yielded the 500~nm $\sim$60~fs pump pulse. The idler output (2073~nm, $\sim$50~fs) of the other optical parametric amplifier was split in a 9:1 ratio. The more intense portion was used as the push, and the weaker portion was used as the probe. The pump and probe beams were collinearly overlapped and focused onto a $\sim$0.2~mm diameter spot on the sample. The push was focused less tightly ($\sim$0.4~mm) to facilitate beam overlap and focused in the same spot in non-collinear ($\sim$5$^{\circ}$) geometry with respect to the pump/probe. The transmission of the probe through the sample was recorded with an amplified PbSe photodetector (PDA20H-EC, ThorLabs). This was connected to a lock-in amplifier (SR830, Stanford Research Systems) coupled to a chopper in the pump beam. The chopper operated at half the repetition rate of the amplifier to block every other pump pulse. The delay between the pump, push and probe beams were controlled with mechanical delay stages. The sample was kept in a nitrogen-purged quartz cuvette during measurements.

\section{\label{sec:level3}Experimental results}
\begin{figure}[h]
\includegraphics[width=7cm]{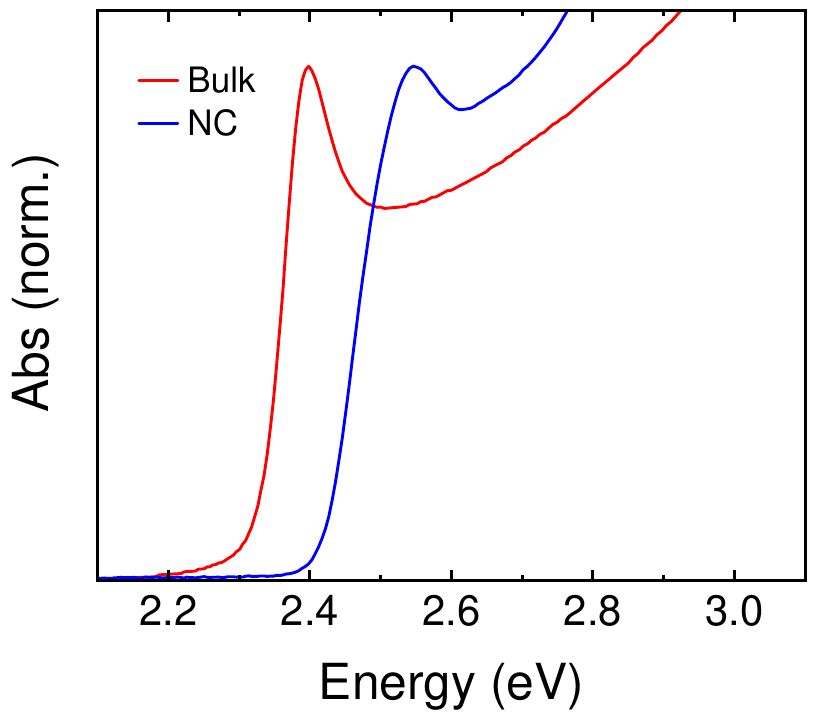}
\caption{\label{Fig. 1}UV-Vis absorption spectra of bulk and NC CsPbBr\textsubscript{3}.}
\end{figure}

The linear absorption spectra of the bulk and NC samples is shown in \cref{Fig. 1}. As expected, the NC sample exhibits a slightly blueshifted absorption onset due to the effect of weak quantum confinement.\cite{Protesescu2015} The NCs studied here have a cubic shape and an average edge-length of 8$\pm$2~nm, as confirmed by transmission electron microscopy (Fig. S1).
\begin{figure*}[ht!]
\includegraphics{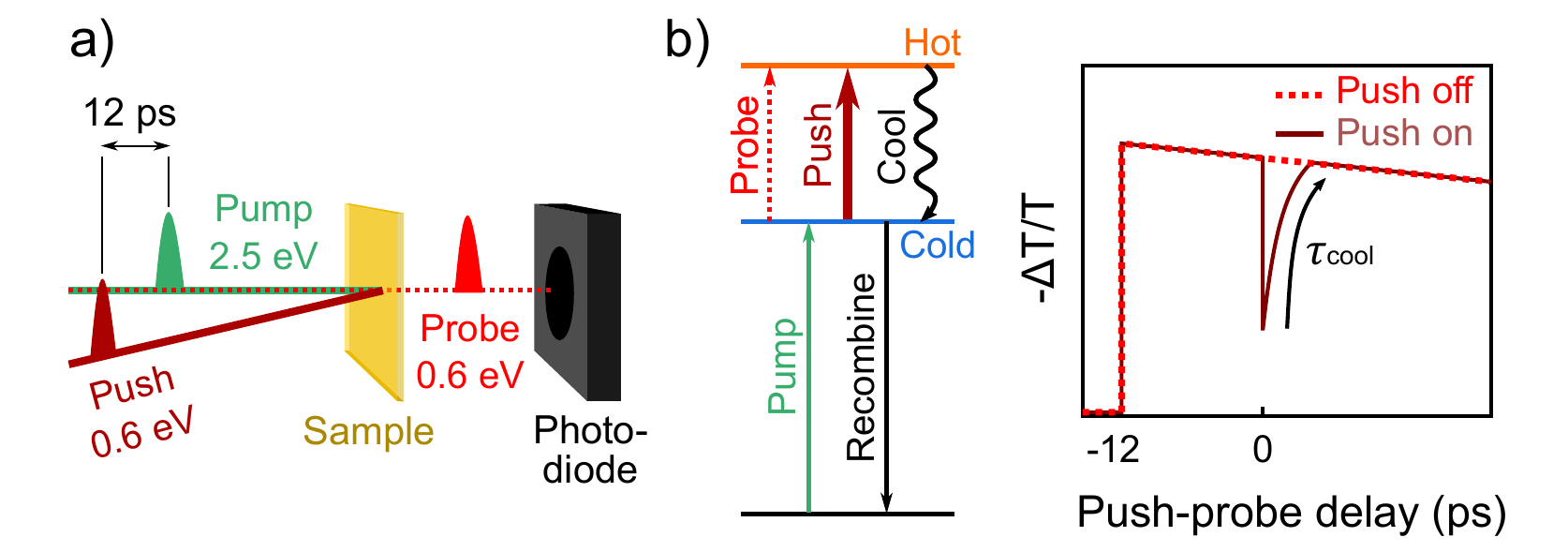}
\caption{\label{Fig. 2}(a) Schematic for the pump-push-probe experiment. Note that the pump and probe are co-linear. (b) Three-level model (left) used to interpret the differential probe transmission in response to the pump and push pulses (right).}
\end{figure*}

To investigate the intraband cooling dynamics in the samples, we used a “visible pump-IR push-IR probe” pulse sequence, which has been outlined in previous works.\cite{Guyot-Sionnest1999,Rabouw2015a,Hopper2018} As depicted in \cref{Fig. 2}, a 2.5~eV pump pulse generates excited carriers which subsequently relax to the band edges. These free carriers are monitored by the differential transmission of a near-IR probe at 0.6~eV, which corresponds to an intraband absorption in the LHPs.\cite{Munson2017} After a fixed delay (12~ps), an intense 0.6~eV push pulse re-excites these cold states, thus transforming them to hot carriers. The heating of the cold carriers is observed as a bleaching of the differential transmission signal. As the hot carriers cool, the pump-probe signal recovers. We fit the recovery of the probe signal with a monoexponential curve to extract the cooling time, $\tau_{cool}$, for the hot carriers. The intensity of the pump and push pulses are used to control the cold and hot carrier densities, respectively. Fluence-dependent pump-probe kinetics for the NC sample are shown in Fig. S2.

The total number of (hot and cold) carriers per unit volume in the system, $n_{exc}$, is calculated using the \cref{eq 1} below. In this equation, $E_{inc}$ is the incident energy per pump pulse, $E_{exc}$ is the energy per photon, $A_I$ is the absorbance of the sample at the pump wavelength, $R$ is the beam radius and $d$ is the sample thickness. The initial hot carrier density, $n_{\sm{0}}^{hot}$, can be obtained from the product of $n_{exc}$ and the ratio between the amplitude of the push-induced bleach to the amplitude of the signal before the arrival of the push pulse.
\begin{equation}
\label{eq 1}
n\textsubscript{exc} = \dfrac{E\textsubscript{inc}(1-10\textsuperscript{-A\textsubscript{I}})}{\pi\ R\textsuperscript{2}\ d\ E\textsubscript{exc}}\
\end{equation}

\begin{figure}
\includegraphics[width=7cm]{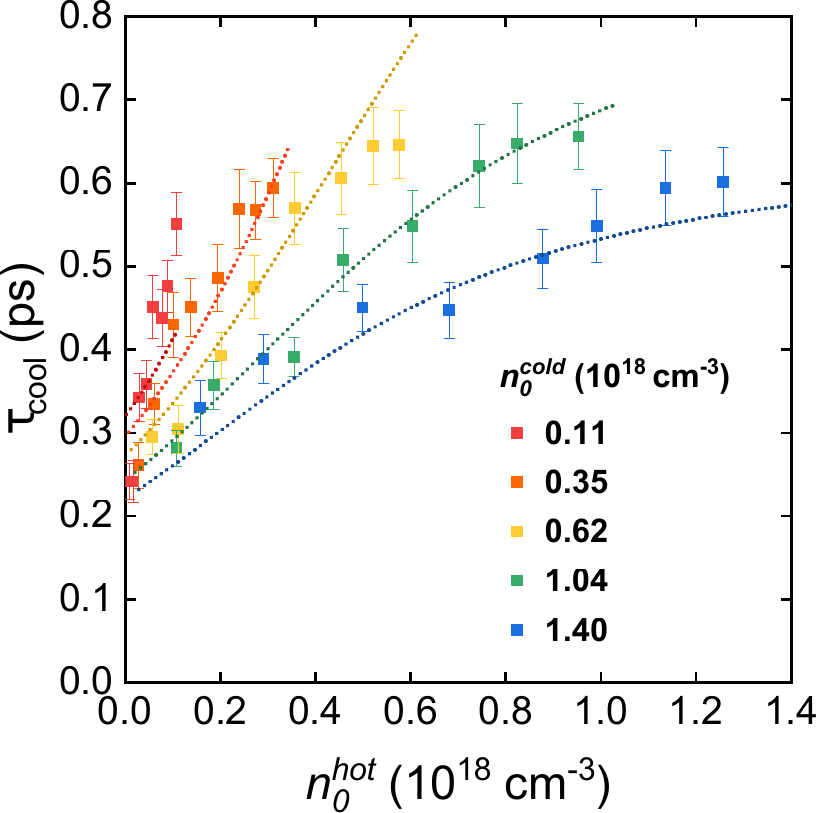}
\caption{\label{Fig. 3}Plot of cooling time with respect to the hot carrier density directly after the push ($n_{\sm{0}}^{hot}$) and cold carrier density just before the push ($n_{\sm{0}}^{cold}$) in the CsPbBr\textsubscript{3} NCs. Square points are experimental data and the dotted lines are fits generated from the model described in \cref{sec:level4}.}
\end{figure}

The dependence of $\tau_{cool}$ on the initial density of the hot and cold carriers is outlined in \cref{Fig. 3}. For a given value of $n_{\sm{0}}^{cold}$ (cold carrier density just before the push), we find that $\tau_{cool}$ increases with increasing $n_{\sm{0}}^{hot}$. This is in accordance with our previous findings, and can be explained by the reduced rate of carrier-phonon interactions through the hot-phonon bottleneck effect.\cite{Hopper2018} Interestingly, we also find that $\tau_{cool}$ tends to decrease with increasing $n_{exc}$, suggesting that the hot-phonon bottleneck effect plays less of a role in intraband cooling when there are many cold carriers in the system. From this, we can deduce that the carrier cooling occurs via at least two competing pathways. When there are few cold carriers, carrier-phonon interactions dominate over the hot-cold carrier interactions, and the hot-phonon bottleneck effect primarily governs the carrier cooling dynamics. At higher cold carrier density, carrier cooling via carrier-carrier interactions starts to dominate, resulting in the overall faster cooling and the reduced impact of the hot-phonon bottleneck effect.

\section{\label{sec:level4}Kinetic modelling}
As discussed above, we interpret our experimental data as the cooling of hot carriers in the LHP samples through carrier-carrier and/or carrier-phonon interactions. To gain more insight into the interplay between these processes, we have developed a numerical model describing each process that is eventually reduced to a single equation, shown later.
To fully describe the data, we take the following approximations:
\begin{itemize}
    \item Carriers can be either hot or cold, hence the sum of the number of hot carriers ($n_{hot}$) and the number of cold carriers ($n_{cold}$) is equal to the total number of carriers ($n_{exc}$); 
    \item Before the push arrives, the number of hot carriers is negligible;
    \item The process of carrier cooling involves two co-existing pathways, each with characteristic rates; (i) carrier-carrier interaction, when a hot carrier scatters with another hot or cold carrier and loses its excess energy; and ii) carrier-phonon interaction, when the excess energy is transferred into an unoccupied phonon mode. The phonon mode becomes occupied, and the carrier cools down;
    \item Total number of phonon modes ($N_{ph}$) available is a constant for the material. Phonon modes can be either occupied or unoccupied.
\end{itemize}

All simulations are presented for the unit volume (cm\textsuperscript{-3}). The initial system of equations is shown in \cref{eq 2} with the initial conditions shown in \cref{eq 3}:  
\begin{equation}
\label{eq 2}
\begin{cases} 
\dfrac{dn_{hot}}{dt} =\tilde{I}_{push}n_{cold}-\alpha n_{exc}n_{hot}-\beta n_{ph}n_{hot}\\
\dfrac{dn_{cold}}{dt}=-\dfrac{dn_{hot}}{dt}-\epsilon n_{cold}\\
\dfrac{dn_{ph}}{dt}=-\beta n_{ph}n_{hot}+\nu(N_{ph}-n_{ph})
\end{cases}
\end{equation}
\begin{equation}
\label{eq 3}
\begin{cases} 
n_{hot}=0\\
n_{cold}=n_{\sm{0}}^{cold}\\
n_{ph}=N_{ph}
\end{cases}
\end{equation}
where $n_{hot}$, $n_{hot}$ and $n_{exc}$ are the number of hot, cold and total number of carriers, respectively; $n_{hot}$, $n_{hot}$ and $n_{exc}$ are the total number of phonon modes and the number of unoccupied phonon modes, respectively; $n_{\sm{0}}^{cold}$ is the cold carrier density at t=0 (just before the push pulse arrives); $\tilde{I}_{push}$ is the number of photons absorbed from the push pulse per unit time, expressed as a time-dependent Gaussian envelope function; $\alpha$ is the rate constant for the carrier-carrier interaction; $\beta$ in the rate constant for carrier-phonon interaction; $\epsilon$ is the rate constant for the recombination of carriers; and $\nu$ is the rate constant for phonon vacancy freeing.

\begin{figure}[hb!]
\includegraphics[width=7cm]{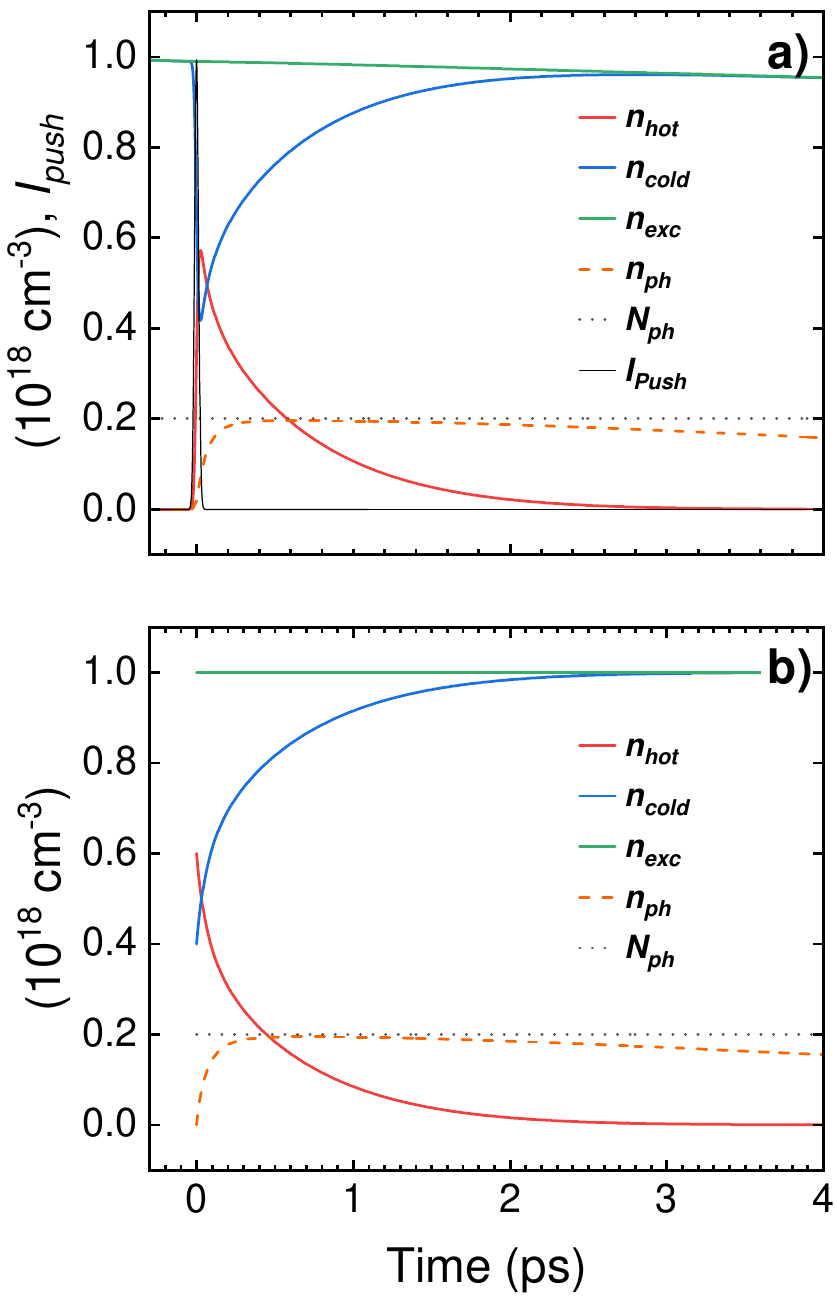}
\caption{\label{Fig. 4} 
(a) Representative plot of carrier dynamics from equations (\cref{eq 2}) and (\cref{eq 3}). Constant parameters used: $\alpha=1.4\times10^{-18}$~cm\textsuperscript{3}~ps\textsuperscript{-1}, $\beta = 2.7\times10^{-17}$~cm\textsuperscript{3}~ps\textsuperscript{-1},  $\epsilon=1.0\times10^{-2}$~ps\textsuperscript{-1},  $\nu = 1.0\times10^{-1}$~cm~ps\textsuperscript{-1}. Boundary conditions used: $n^{cold}_{\sm{0}} = 1.0\times10^{18}$, $n^{hot}_{\sm{0}} = 6.0\times10^{17}$, $N_{ph} = 2.0\times 10^{17}$. Pulse duration: 35 fs. (b) Representative plot of carrier dynamics from equations \cref{eq 4} and \cref{eq 5}, using the same boundary conditions as above. Constant parameters used: $\alpha=1.4\times10^{-18}$~cm\textsuperscript{3}~ps\textsuperscript{-1}, $\beta = 2.7\times10^{-17}$~cm\textsuperscript{3}~ps\textsuperscript{-1}, $\nu = 1.0\times10^{-1}$~ps\textsuperscript{-1}. The same boundary conditions as in \cref{Fig. 4}(a) were used. All parameters are in roughly the same order of magnitude as the experimentally determined values.}
\end{figure}

The hot-phonon bottleneck effect arises from the $-\beta n_{ph}n_{hot}$ component of the rate equation for $n_{hot}$; this component is proportional to the unoccupied phonon density, which depletes during the carrier-phonon interaction. The hot-phonon bottleneck effect becomes prominent if the rate of phonon depletion is greater than or comparable to the rate of phonon scattering (the dissipation of hot phonons to the lattice), i.e. $-\beta n_{ph}n_{hot} \geqslant   \nu (N_{ph}-n_{ph}$). The effect of carrier density is expressed in $-\alpha n_{exc} n_{hot}$; an increased rate of carrier-carrier interactions is expected for a greater number of carriers. This system can be numerically solved to provide a full picture of the process, as shown in \cref{Fig. 4}(a). 
The result is comprehensive but involves too many parameters for the analysis of carrier dynamics. Considering that the push pulse is short ($\sim$50~fs) in comparison to the cooling times ($\sim$500~fs),\cite{Hopper2018} we can assume that the intraband excitation is immediate and hence $\tilde{I}_{push} n_{cold}\approx0$ after 100~fs, i.e. in the time window of our interest. In addition, we notice that the recombination rate is considerably slow (>100~ps),\cite{Makarov2016} and thus can be neglected ($\epsilon\approx0$). Consequently, $n_{exc}$ becomes a constant ($n_{exc} = n_{\sm{0}}^{cold}$) and the rate equation for $n_{cold}$ becomes obsolete. The system simplifies into the form shown in \cref{eq 4}, with the initial conditions shown in \cref{eq 5}:
\begin{equation}
\label{eq 4}
\begin{cases} 
\dfrac{dn_{hot}}{dt} =-\alpha n_{exc}n_{hot}-\beta n_{ph}n_{hot}\\
\dfrac{dn_{ph}}{dt}=-\beta n_{ph}n_{hot}+\nu(N_{ph}-n_{ph})
\end{cases}
\end{equation}
\begin{equation}
\label{eq 5}
\begin{cases} 
n_{hot}=n_{\sm{0}}^{hot}\\
n_{ph}=N_{ph}
\end{cases}
\end{equation}
where $n_{\sm{0}}^{hot}$ is proportional to $\tilde{I}_{push}$. The solution of these equations provides the time dependence of $n_{hot}$ and $n_{ph}$, plotted in \cref{Fig. 4}(b).

The rate equation for $n_{hot}$ involves a variable ($n_{ph}$). This can be eliminated using the following expression from the rate equation of $n_{hot}$ and $n_{ph}$. All other parameters are constants.

\begin{equation}
\label{eq 6}
\dfrac{dn_{hot}}{dn_{ph}} = \dfrac{-\alpha n_{exc}n_{hot}-\beta n_{ph}n_{hot}}{-\beta n_{hot} n_{ph}+\nu(N_{ph}-n_{ph})}\approx \dfrac {\alpha n_{exc}+\beta n_{ph}}{\beta n_{ph}}
\end{equation}

We can neglect the $\nu(N_{ph}-n_{ph})$ term based on the assumption that the phonon scattering is significantly slower than the carrier-phonon interaction. This notion is supported by theoretical and experimental findings of >1~ps phonon lifetimes in LHPs.\cite{Wang2016c,Whalley2016,Beecher2016,Li2017a}

Using the initial condition shown in \cref{eq 5}:
\begin{equation}
\label{eq 7}
\int\limits_{{n_{\sm{0}}^{hot}}}^{n_{hot}} dn^{*}_{hot}= \int\limits_{N_{ph}}^{n_{ph}}\frac{\alpha n_{exc}+\beta n_{ph}^*}{\beta n_{ph}^*}dn_{ph}^*
\end{equation}

This simplifies to:
\begin{equation}
\label{eq 8}
n_{ph}=N_{ph}\exp\bigg(\frac{\beta}{\alpha n_{exc}} (\Delta n_{hot}- \Delta n_{ph})\bigg)
\end{equation}

where $\Delta n_{hot}=n_{hot}-n_{\sm{0}}^{hot}$ and $\Delta n_{ph}=n_{ph}-N_{ph}$.\\

We use an alternative approach to derive the equation for further simplification and to find a physical quantity related to polaron overlap. Let’s assume that each hot carrier occupies a specific volume, $V_p$, where carrier-phonon coupling can occur.\cite{Frost2017a} Then we divide a unit volume (V) of the lattice into uniform cells with the volume $V_p$. The hot carriers can occupy any of the cells and can either overlap or not overlap with each other. The phonons are assumed to be evenly scattered across the lattice, and only one of the overlapping hot carriers can interact with the phonon modes in the cell. When $(n^{hot}_{\sm{0}} - n_{hot})V$ carriers are coupled and dropped into the cells with the volume $V_p$ in a lattice with $N_{ph}$ phonons per unit volume, then the average available number of accessible phonons per unit volume ($n_{ph}$) can be expressed as follows:
\begin{equation}
\label{eq 9}
n_{ph}= N_{ph}\bigg(1-\frac{V_p}{V}\bigg)^{(n_{\sm{0}}^{hot}-n_{hot})V}
\end{equation}

Given than $\frac{V_p}{V}$ is a small number, the expression is approximated to the following:

\begin{equation}
\label{eq 10}
n_{ph} = N_{ph}\exp(V_p \Delta n_{hot})
\end{equation}

\begin{figure}[h]
\includegraphics[width=7cm]{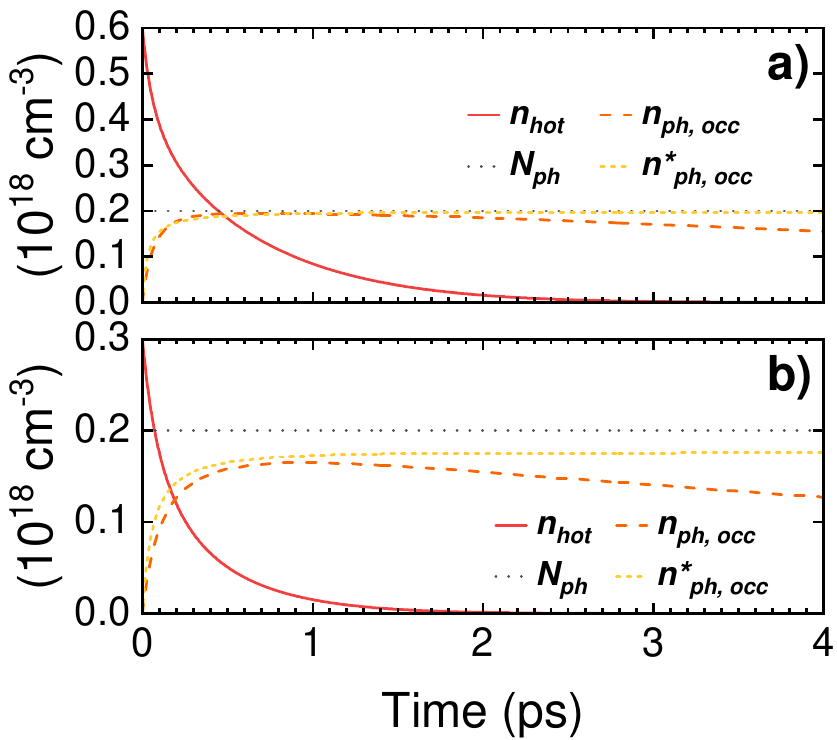}
\caption{\label{Fig. 5} 
The dynamic occupied phonon density $n_{ph, occ} = N_{ph}-n_{ph}$   from equations (\cref{eq 4}) and (\cref{eq 5}) is shown as the orange dashed line, and the static occupied phonon density $n^*_{ph, occ}$ from (\cref{eq 10}) is shown as the yellow dashed line. (a) and (b) show the effect when the initial hot carrier density is set to $6.0\times10^{17}$ and $3.0\times10^{17}$ cm\textsuperscript{-3}, respectively. All other parameters are kept constant: $\alpha=1.4\times10^{-18}$ cm\textsuperscript{3} ps\textsuperscript{-1}, $\beta = 2.7\times10^{-17}$ cm\textsuperscript{3} ps\textsuperscript{-1},  $\nu = 1.0\times10^{-1}$ ps\textsuperscript{-1}, $n_{exc} = 1.0\times10^{18}$ cm\textsuperscript{3}, $V_p=7.0\times10^{-18}$ cm\textsuperscript{3}. Boundary condition: $N_{ph}=2.0\times10^{17}$ cm\textsuperscript{-3}.}
\end{figure}

\cref{Fig. 5} (a) and (b) compare the expression of $n_{ph}$ from \cref{eq 4,eq 5,eq 10} in the time domain. We demonstrate that the use of \cref{eq 10} is a reasonable approximation of $n_{ph}$ in equations (4) and (5), especially at early times <1 ps.
\newpage

The rate equation of $n_{hot}$ can be expressed as: 
\begin{equation}
\label{eq 11}
\dfrac{dn_{hot}}{dt}=-\alpha n_{exc}n_{hot}-\beta N_{ph} \exp(-V_p n^{hot}_{\sm{0}})n_{hot}
\end{equation}

where $\alpha$, $\beta$, $V_p$ and $N_{ph}$ are constants. In this expression, $\alpha$ is the rate constant for the carrier-carrier interaction, $\beta$ is the rate constant for the carrier-phonon interaction, $V_p$ is the volume of a polaron and $N_ph$ is the number of unoccupied phonon modes per unit volume in the lattice.

The equation \cref{eq 11} includes the product of two arbitrary constants ($\beta N_{ph}$), allowing the elimination of a parameter. $\beta$ is defined as the average volume of space around a hot carrier where a phonon mode will interact with the carrier with $\sim$100\% probability in 1 ps. $\beta N_{ph}$ can be substituted by $\rho V_p$ where $V_p$ is the volume of space around a hot carrier where phonons are coupled and $\rho$ is the number of carrier-phonon interactions per unit volume per 1 ps. Both $\beta N_{ph}$ and $\rho V_p$ represent the number of carrier-phonon interactions per hot carrier per 1 ps. Hence, the rate equation can be expressed as the \cref{eq 12}.
\begin{equation}
\label{eq 12}
\dfrac{dn_{hot}}{dt}=-\alpha n_{exc}n_{hot}-\rho V_p \exp(-V_p n^{hot}_{\sm{0}})n_{hot}
\end{equation}

The hot-phonon bottleneck arises from the reduced number of available phonon modes with the increasing hot carrier density, expressed as $\rho V_p \exp(-V_p n^{hot}_{\sm{0}})$. The carrier-carrier interaction is shown in the $-\alpha n_{exc} n_{hot}$ component; the rate of the interaction is shown to be proportional to the carrier density. The cooling time is calculated using the following expression:
\begin{equation}
\label{eq 13}
\tau_{cool} = \int\limits_{n_{\sm{0}}^{hot}}^{\frac{n^{hot}_{\sm{0}}}{e}}\dfrac{dn_{hot}}{dt}dt
\end{equation}

\begin{figure}[h]
\includegraphics[width=7cm]{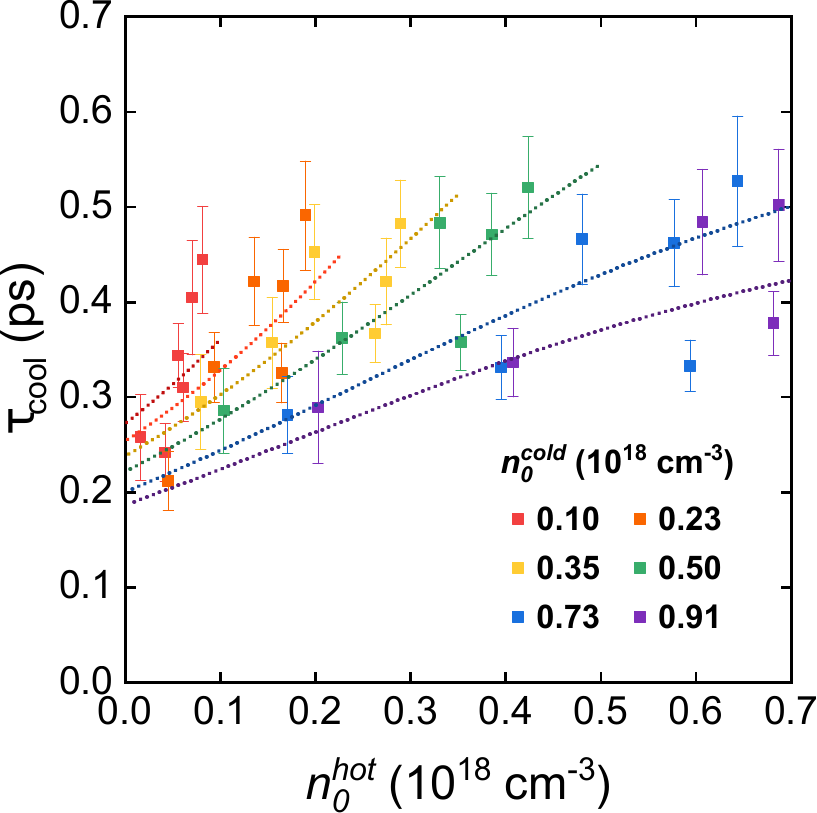}
\caption{\label{Fig. 6} Plot of experimental data (square points) and fits from the model (dashed lines) for bulk CsPbBr\textsubscript{3}.}
\end{figure}

Based on this equation, we fit the experimental data from the CsPbBr\textsubscript{3} NCs and bulk material for comparison. The fits and experimental data are shown in \cref{Fig. 3,Fig. 6} for the respective samples.  Material constants derived from the fits are tabulated in \cref{tab:table1}.

\begin{table}[h]
\caption{\label{tab:table1}Material constants for the polaron volume ($V_p$), carrier-carrier interaction ($\alpha$) and carrier-phonon interaction ($\rho$) derived from the fit.}
\begin{ruledtabular}
\begin{tabular}{cccc}
Material&$V_p$&$\alpha$&$\rho$\\
~&$(10^{-18} cm^3)$&$(10^{-18} cm^3 ps^{-1})$&$(10^{18} cm^{-3} ps^{-1})$\\
\hline
Bulk CsPbBr\textsubscript{3}&7.58&1.95&0.46\\
CsPbBr\textsubscript{3} NCs&7.00&1.08&0.43\\
\end{tabular}
\end{ruledtabular}
\end{table}

Based on $V_p$, the polaron radius in bulk CsPbBr\textsubscript{3} and the CsPbBr\textsubscript{3} NCs is 12.1 and is 11.9~nm, respectively. Although “large” polarons are often invoked to explain the unique photophysics of LHPs,\cite{Zhu2015,Zhu2016,Miyata2017,Wang2018,Batignani2018,Lan2018,Cinquanta2018,Mahata2019} we note these values are approximately double to the polaron size experimentally determined by Munson et al. for MAPbI\textsubscript{3} ($\sim$4.5~nm radius),\cite{Munson2018} and larger still for the computationally-derived values for CsPbBr\textsubscript{3} reported elsewhere.\cite{Miyata2017,Hopper2018} We posit that the relatively large values obtained here are a consequence of energy transfer processes which lead to an incorrect estimation of the local carrier density. In other words, we believe that energy transfer processes result in “hot spots” where the local carrier density is higher than the average carrier density of the photoexcited ensemble. Experimental evidence for this phenomenon has been reported in LHPs.\cite{Vrucinic2015} We also note that the comparable values for the bulk and NC samples are not entirely unexpected given the bulk-like behaviour of weakly confined CsPbBr\textsubscript{3} NCs.~\cite{Butkus2017}

Taking that both the carrier-carrier and carrier-phonon interactions occur on the sub-ps timescale, we can use the model parameters to compare the carrier-carrier and carrier-phonon interaction distances.  The values above indicate that for CsPbBr\textsubscript{3} and its NC analogue, the typical distance at which carrier-phonon interactions start playing a role is $\sim$12~nm. This is slightly larger than the typical carrier-carrier interaction distance which, when deduced from $\alpha$, is equal to $\sim$7~nm. As discussed above, these absolute values are higher estimates of the actual numbers due to aforementioned incorrect estimation of the local carrier density. However, the estimates of $\alpha$ and $V_p$ depend on carrier density in the same way, so the ratio between these parameters should be captured by the model correctly. The fact that the carrier-carrier interactions are active at shorter distances than the polaron size is consistent with the notion of polaronic screening in LHPs previously reported in the literature.\cite{Zhu2015,Zhu2016,Miyata2017}

\section{Conclusion}
In summary, we developed a numerical model which includes “carrier-carrier” and “carrier-phonon” interaction terms to describe the role of the hot and cold carriers in the carrier cooling kinetics of bulk and nanocrystal CsPbBr\textsubscript{3} films determined by “pump-push-probe” spectroscopy. The observation of slower cooling kinetics at higher push intensity (higher hot carrier density) is explained by invoking a hot-phonon bottleneck mechanism where overlapping hot polarons compete for phonons in order to cool. This is incorporated in our model by the carrier-phonon term which describes the spatial overlap between the hot polaron states. Meanwhile, the observation of faster cooling kinetics for higher pump intensity is attributed to the redistribution of energy from a hot carrier to adjacent cold carriers, as described by the carrier-carrier term. By fitting our model to the experimental results, we demonstrate the interplay between these contrasting effects, and propose a way to estimate the polaron size, the phonon density of the lattice and the rate constant for carrier-carrier and carrier-phonon interactions. 

\begin{acknowledgments}
A.A.B. is a Royal Society University Research Fellow. A.J. thanks the Royal Society for supporting the project. A.G. thanks Mr. Amirlan Sekenbayev of Queen Mary University for useful discussions. This project has also received funding from the European Research Council (ERC) under the European Union’s Horizon 2020 research and innovation programme (Grant Agreement No. 639750).
The authors declare no competing financial interest. 
\end{acknowledgments}

\bibliography{JCPPaper}

\end{document}